\documentclass[fleqn, usenatbib]{mnras}

\usepackage[T1]{fontenc}
\usepackage[utf8]{inputenc}
\usepackage{lmodern}
\usepackage{graphicx}	
\usepackage{amsmath}	
\usepackage{amssymb}	
\usepackage{multicol}        
\usepackage{bm}		
\usepackage{pdflscape}	
\usepackage{cuted} 
\usepackage{color}
\usepackage{array}

\graphicspath{{./figs/}}

\defcitealias{Rodrigues:2018duc}{R18}

\def\d{{\rm d}}
\DeclareMathOperator\erf{erf}

\newcommand{\mscript}[1]{{\mbox{\scriptsize #1}}}
\newcommand{\mtiny}[1]{{\mbox{\tiny #1}}}
\newcommand{\Rteen}{\citetalias{Rodrigues:2018duc}}
\newcommand{\bulge}{\mscript{b}}
\newcommand{\disk}{\mscript{d}}
\newcommand{\gas}{\mscript{gas}}
\newcommand{\Newt}{\mscript{N}}
\newcommand{\Circ}{\mscript{C}}
\newcommand{\model}{\mscript{M}}
\newcommand{\loga}{{A_0}}
\newcommand{\logyb}{{Y_\bulge}}
\newcommand{\logyd}{{Y_\disk}}

\definecolor{valecol}{rgb}{0,0.5, 1.}
\definecolor{davicol}{rgb}{0.7,0.3, 0.5}

\newcommand{\natastron}{{Nat. Astron.}}

\title[A fundamental test for MOND]{A fundamental test for MOND}

\author[Marra, Rodrigues \& de Almeida]{ 
Valerio Marra$^{1}$\thanks{E-mail: marra@cosmo-ufes.org},
Davi C. Rodrigues$^{1}$\thanks{E-mail: davi.rodrigues@cosmo-ufes.org}
and Álefe O. F. de Almeida$^{1,2}$\thanks{E-mail: alefe@ibm.com}\\
$^1$ Núcleo de Astrofísica e Cosmologia, PPGCosmo \& Dep.~de Física, Universidade Federal do Espírito Santo, 29075-910, ES, Brazil\\
$^2$ IBM Brasil,  29055-131, Vitória, ES, Brazil.
}

\date{Accepted XXX. Received YYY; in original form ZZZ}

\pubyear{2020}

\begin{document}
\label{firstpage}
\pagerange{\pageref{firstpage}--\pageref{lastpage}}

\maketitle

\begin{abstract}
The Radial Acceleration Relation (RAR) shows a strong correlation between two accelerations associated to galaxy rotation curves. The relation between these accelerations is given by a nonlinear function which depends on an acceleration scale $a_\dagger$. Some have interpreted this as an evidence for a gravity model, such as Modified Newtonian Dynamics (MOND), which posits a fundamental acceleration scale $a_0$ common to all the galaxies.
However, it was later shown, using Bayesian inference, that this seems  not to be the case: the $a_0$ credible intervals for individual galaxies were not found to be compatible among themselves. This type of test is a fundamental test for MOND as a  theory for gravity, since it directly evaluates its basic assumption and this using the data that most favor MOND: galaxy rotation curves. 
Here we improve upon the previous analyses by introducing a  more robust method to assess the compatibility between the credible intervals, in particular without Gaussian approximations. We  directly estimate, using a Monte Carlo simulation, that the existence of a fundamental acceleration is incompatible with the data at more than $5\sigma$. We also consider quality cuts in order to show that our results are robust against outliers. In conclusion, the new analysis further supports the claim that the acceleration scale found in the RAR is an emergent quantity.
\end{abstract}


\bigskip

\section{Introduction}\label{intro}

Several and diverse independent observations -- spanning a large range of scales in space and time -- strongly suggest the existence of some type of dark matter. The standard Cold Dark Matter (CDM) paradigm has been successful in providing relevant predictions and important insights. However, direct detection efforts are still inconclusive and many different dark matter candidates agree with the current observational and experimental bounds \citep[e.g.,][]{0521857937, 9781786340016}. Consequently, there is a large level of uncertainty as far as the status of the nature of dark matter is concerned.

The Radial Acceleration Relation (RAR) \citep{McGaugh:2016leg}, which is  closely related to the Mass Discrepancy-Acceleration Relation \citep{1990A&ARv...2....1S, McGaugh:2004aw, Milgrom:2016uye},  shows a tight correlation between two accelerations associated to galaxy rotation curves: one is computed from the observed redshift and the other is the expected gravitational acceleration due to baryonic matter alone. The relation between these accelerations is given by a nonlinear function that depends on an acceleration scale, labeled $a_\dagger$. \citet{Milgrom:2016uye} and \citet{Li:2018tdo} have interpreted the appearance of this scale as  evidence against CDM and in favor of the Modified Newtonian Dynamics (MOND) model that depends on a fundamental acceleration scale $a_0$, which numerically  would be close to the value of $a_\dagger$. On the other hand, different results within CDM are also capable of explaining, at least in part, the emergence of RAR \citep{Ludlow:2016qzh, Navarro:2016bfs, 2018MNRAS.476.3816F, Dutton:2019gor}. In particular, the tightness of the RAR, i.e.~its dispersion, does not seem to be at odds with the CDM paradigm \citep{2019ApJ...882....6S}.

\citet[][hereafter R18]{Rodrigues:2018duc} have shown%
\footnote{See also \cite{McGaugh:2018aa, Kroupa:2018kgv, Rodrigues:2018lvw, 2020NatAs.tmp...15C, Rodrigues:2020squ}.
For a different approach and with the same data, see also \citet[][]{Frandsen:2018ftj, Zhou:2020tst}.} that the RAR, rather than suggesting a 
fundamental $a_0$, actually provides strong evidence against the existence of such fundamental constant, implying that the $a_\dagger$ scale must be emergent (among other examples of an emergent scale in the galaxy context, we recall the disk scale length). \citetalias{Rodrigues:2018duc} was the first work that, in order to conclude on the universality of the acceleration scale, studied the (Bayesian) posterior distributions on the acceleration scales inferred from individual galaxies from a large dataset. For previous studies with error bars, usually defined from a given change on the $\chi^2$ value with respect to its minimum, see \citet{Randriamampandry:2014eoa} and references therein.
The result by \citetalias{Rodrigues:2018duc} was subsequently confirmed by \citet{Chang:2018lab}, which repeated the \Rteen{} analysis considering the priors of \citet{Li:2018tdo}. They directly confirmed that the latter priors   also reject the fundamental acceleration hypothesis with high confidence (clearly beyond the 5$\sigma$ level), these results are also in agreement with \citetalias{Rodrigues:2018duc} and \cite{Rodrigues:2018lvw}. More recently, \cite{Zhou:2020tst} have also provided further support.

Here, we improve upon the analysis of \Rteen{}. First, we use the same priors and nuisance  parameters that were used to support the existence of a fundamental $a_0$ in \citet{McGaugh:2016leg} and \citet{Li:2018tdo}. This set of priors  is a physically reasonable choice, it was used in different works and it was shown to generate $a_0$ credible intervals for individual galaxies that are larger (and so more conservative) than those obtained with the set of priors adopted in \Rteen{} \citep[see][for further details]{Rodrigues:2018lvw}.
Second, and this is the main novelty of the present work, we go beyond the approximations that \Rteen{} adopted in order to quantify the tension between the posteriors on $a_0$ of the SPARC galaxies. We achieve the latter via an extension of the  \cite{Verde:2013wza} proposal. Our new results are not based on Gaussian approximations in order to compare the posteriors. With the new and more accurate method, we confirm here the findings of \Rteen{} that a fundamental acceleration scale is incompatible with  rotation curve data. We add that the method here proposed is not specific to galaxy data and should be useful in other contexts as well.

This paper is organized as follows.
In Section~\ref{mond} we review the theory behind the fundamental acceleration scale $a_0$, while in Section~\ref{se3} we discuss the observational data.
In Section~\ref{analysis} we describe how we obtain and compare the posterior distributions on $a_0$,  in Section~\ref{NumMethods} we summarize our numerical methods and our results are presented in Section~\ref{results}. Section~\ref{conclusions} is devoted to our conclusions.
Technical details are given in Appendixes~\ref{app:mcmc},~\ref{app:ext} and ~\ref{app:revGau}.

\section{The RAR and MOND}\label{mond}

\subsection{RAR for individual galaxies}

Since the RAR is a tight correlation for rotationally supported galaxies \citep{McGaugh:2016leg} (tight in the sense that it spans about four orders of magnitude on the baryonic acceleration with a rms dispersion of about 0.1 dex), it is relevant to ask whether the $a_\dagger$ scale inferred from the RAR could be a universal scale for individual galaxies. In principle, this could unveil unexpected baryonic or dark matter properties. As shown by \cite{Li:2018tdo}, for this same sample of galaxies, assuming that the RAR is relevant for individual galaxies with a common $a_0$ value given by $a_\dagger$, it is possible to infer a tighter correlation between the baryonic and observed accelerations. The latter is achieved if one considers best fits for the  observed acceleration data and  considering mass-to-light ratios, distance and inclination as nuisance parameters whose priors are compatible with the observational errors stated by \citet{2016AJ....152..157L}.
From the assumption that the RAR is valid for individual galaxies, the latter analysis is capable of finding a residual scatter in the data which must be attributed to observational error, according the the starting hypothesis. However, this approach is not capable of concluding if the data is compatible with the assumption that the RAR is valid for individual galaxies. There is no logical necessity that the resulting minimization procedure, in spite of minimizing the scatter, would lead to a picture closer to reality than the original RAR approach (the use of observational errors to decrease a scatter may be an artificial procedure). Indeed, similar criticism was brought forward  by \cite{Dutton:2019gor, 2019ApJ...882....6S}.

\subsection{MOND}

Closely related to the hypothesis of the RAR being valid for individual galaxies is the Modified Newtonian Dynamics (MOND) hypothesis \citep{1983ApJ...270..371M, Famaey:2011kh}. MOND assumes no dark matter in galaxies and is based on a non-linear relation between the physical acceleration ($\boldsymbol a$) and the Newtonian one ($\boldsymbol{a}_\mscript{N}$), written as
\begin{equation}
	\boldsymbol{a} = \nu \Big ( \frac{a_\mscript{N}}{a_0}\Big) \boldsymbol{a}_\mscript{N} \,,
\end{equation}
where $\nu$ is a function, commonly called the (inverse) interpolating function of MOND, and $a_0$  would be a  fundamental constant with dimension of acceleration. The interpolating function needs to be such that for large accelerations ($a_\mscript{N} \gg a_0$) one finds standard Newtonian gravity (i.e., $\boldmath{a} = \boldmath{a}_\mscript{N}$), while for small accelerations ($a_\mscript{N} \ll a_0$) one finds the so-called deep-MOND regime. The deep-MOND regime is taken to be $a = \sqrt{a_0 a_\mscript{N}}$, which implies that far from the object of mass $M$ the circular velocity of a test particle is independent on its distance from $M$: $V^2 = \sqrt {a_0 G M}$, where $G$ is the (Newtonian) gravitational constant. The latter  behavior was originally motivated from the Tully-Fisher relation \citep{1983ApJ...270..371M}, while more recently other related motivations were put forward  \citep[e.g.,][]{Milgrom:2015ema}.

\subsection{A fundamental test of MOND} \label{FundTest}

In order to apply MOND to individual galaxies, the first step is to select a $\nu$ function and a value for the acceleration scale $a_0$. Different approaches can be found in the literature \citep{1983ApJ...270..371M, Famaey:2011kh}, each one with its pros and cons. The most common approach is to choose a simple function that satisfies the asymptotic properties for being the interpolating function, and then perform rotation curve fits for several galaxies with $a_0$ taken as a free parameter. The global value of $a_0$ is then taken to be certain average (such as the median) over all the best-fit $a_{0 g}$ values (where the index $g$ was added to indicate the value of $a_0$ for the $g$-th galaxy). Assuming that the galaxy sample is a representative one, the resulting global $a_0$ value depends on the adopted interpolating function $\nu$ and on how the uncertainties of the baryonic parameters  were handled.
In the end, with reasonable considerations, the result is  $a_0 \sim 10^{-13} \, \mbox{km/s}^2$.
See, for instance, \Rteen{} (supplementary material) and \citealt{Gentile:2010xt}.

For a given $\nu$ function, the determination of the global $a_0$ value from such procedure is not optimal: it neglects the information from the individual $a_{0 g}$ uncertainties, i.e., the $a_{0 g}$ posterior distributions. 
Actually, not only one has the chance of finding the most accurate global $a_0$ value, from these posteriors one can also perform  {\it the most fundamental test of MOND for a given interpolating function, namely}: {\it to analyze the compatibility of the existence of a fundamental acceleration scale with the observational data.}

Before concluding this subsection, we remark that the issue of constancy of $a_{0}$ among different galaxies is not a new one. In particular, \citet{Kent:1987zz} singled out a factor 5 discrepancy between the best-fit $a_{0 g}$ values from different galaxies, which lead \citet{1988ApJ...333..689M} to point out possible physical issues as the cause of such discrepancy. \citet{1988ApJ...333..689M} also claimed that using  $a_0$ as a free parameter should be deprecated, apart from the purpose of determining the best overall $a_0$ value. Perhaps, at the time,  this research field was not yet ready for such tests (e.g., due to Bayesian methods being uncommon  and due to the lack of computational power), but currently we understand that there is no justification to avoid  this fundamental test. We also point out that in \Rteen{} and here we find a discrepancy of about 2 orders of magnitude among the best-fit $a_{0 g}$ values from individual galaxies.

\subsection{RAR-inspired interpolating function}

The RAR itself provides a data-driven choice for the interpolating function. If MOND is true,  the most sensible way of specifying an interpolating function at galaxy level seems to be looking for RAR-like data, which already displays a correlation between baryonic and observational acceleration with a minimum set of assumptions on the observational data. The most simple and precise analytical function currently known  capable of describing the correlation is given by \citep{McGaugh:2016leg}
\begin{equation} \label{int}
\boldsymbol{a} = \frac{\boldsymbol{a_{\mscript{N}}}}{1 - e^{-\sqrt{a_\mtiny{N}/a_0}}} \,. 
\end{equation}
Here we adopt this interpolating function, which we call the RAR-inspired interpolating function, but with an {\it a priori} unknown value for $a_0$. We do not assume that the acceleration scale is the  one found from the RAR ($a_\dagger$), thus in the above we use $a_0$.  In \Rteen{} we considered two other interpolating functions commonly used in the MOND context, but our results are essentially the same for any of them (the RAR-inspired interpolating function \eqref{int} has a slightly smaller RAR dispersion and fares slightly better with respect to a common acceleration scale).

\section{Observational data modeling} \label{se3}

Here we use the SPARC data \citep{2016AJ....152..157L} for rotationally supported galaxies, which were the same data used to derive the RAR \citep{McGaugh:2016leg}. The same quality cuts applied to the original 175 SPARC galaxies are also applied here, namely that galaxies with inclinations smaller than 30$^\circ$ and those with relevant asymmetries are not considered (i.e., those with quality flag {\cal Q}=3). This leads to a sample of 153 galaxies, which we call the RAR sample.

Since dark matter is not being considered, the Newtonian acceleration only has  the baryonic component, which is subdivided into two main parts:  stellar and atomic gas components (the latter composed by hydrogen and helium mainly; while the former is decomposed into a disk and a bulge components). The centripetal Newtonian acceleration, as inferred from the baryonic distribution, can be decomposed as 
\begin{equation} \label{newtdecompA}
a_\Newt 
= \Upsilon_\bulge a_\bulge + \Upsilon_\disk a_\disk+ a_\gas \,.
\end{equation}
Where $a_\bulge$ and $a_\disk$ refer to the bulge and disk contributions to the centripetal acceleration for mass-to-light ratios ($\Upsilon_\bulge$ and $\Upsilon_\disk$) equal to one. Equivalently, but closer to the provided SPARC data,
\begin{equation} \label{newtdecompV}
V_\Newt^2  
= \Upsilon_\bulge|V_\bulge| V_\bulge + \Upsilon_\disk|V_\disk| V_\disk + |V_\gas| V_\gas \,.
\end{equation}
In the above, we use $V_x |V_x|$ in place of $V_x^2$ since it is customary (and the SPARC database uses this convention) to use negative values of $V_x$ to represent negative contributions to the centripetal acceleration; thus, $V_x(R) <0$ means that $a_x <0$, implying that one should write $a_x = V_x |V_x|/R$. If, for a given $R$, $V_\Newt^2 <0$,  then there is no Newtonian rotation curve at that radius.

Since MOND effectively amplifies the acceleration inferred from the baryons,  uncertainties on the baryonic data have larger impact than in dark matter models. The baryonic parameters whose uncertainties have larger dynamical effect on the inferred circular velocity from MOND are commonly taken to be the stellar mass-to-light ratios ($\Upsilon_\disk$, $\Upsilon_\bulge$), galaxy distance ($D$) and lastly galaxy inclination ($I$) \citep[e.g.,][]{Gentile:2010xt, McGaugh:2016leg}. We remark that \Rteen{} considered $I$ as fixed, since this additional parameter would not impact the conclusions in that paper. There is also an additional technical reason for \Rteen{} to be especially economic on the number of nuisance parameters:  \Rteen{} explored the Bayesian posteriors on a grid, which is a very robust way of sampling the tails of the distributions ($>3\sigma$ credible regions).
However, the discretization of the parameter space is computationally demanding, and any new parameter would at least multiply the necessary computational time by a factor $\mathcal{O}(100)$. Hence, for a larger number of parameters, MCMC methods become necessary and we apply them here (with a lot of care on convergence issues).

The dynamical impact of changes on $\Upsilon$ is already explicit in eq.~\eqref{newtdecompV}, which should be combined with eq.~\eqref{int}. A distance change from $D_0$ to $D$ due to a factor $\delta_D$ (i.e., $D = \delta_D D_0$) implies that: $i$) the galaxy coordinate radius is stretched by $\delta_D$  ($R' = \delta_D R$); $ii$) the luminosity and hence the mass are increased such that the Newtonian acceleration is invariant, and $iii$) there are no changes on the galaxy spectra, hence on the observational circular velocity. In summary (see e.g., supplementary material of \Rteen{}  and \citealt{Li:2018tdo} for further details), 
\begin{equation} \label{Dcorrection}
		V'_\Newt(R') = \sqrt{\delta_D} V_\Newt(R) \; \mbox{ and } V'_\Circ(R')= V_\Circ(R)\,.
\end{equation}
Distance changes do not change the circular velocity $V_\Circ$ (in the sense above), or its errors. It is common to designate $V_\Circ$ as the observed velocity (``$V_{\mscript{Obs}}$''). However, since we are here considering inclination changes, which change the value of the latter,  $V_\Circ$ seems to be a good choice.

The circular velocity depends on the redshift data and on the galaxy inclination (since the redshift only gives line-of-sight velocity information). An inclination change from $I_0$ to $I$  changes  $V_\Circ$ and the corresponding error $\sigma_V$, which are given by \citep[e.g.,][]{1998ApJ...508..132D, 2016AJ....152..157L, Li:2018tdo}
\begin{equation} \label{Icorrection}
	V'_\Circ = V_\Circ \frac{\sin I_0}{\sin I} \; \mbox{ and } \sigma_V' = \sigma_V \frac{\sin I_0}{\sin I} \, .
\end{equation}

\section{Statistical analysis}\label{analysis}

We will carry out Bayesian inference for the galaxies of the RAR sample \citep{2016AJ....152..157L, McGaugh:2016leg}. The posterior distribution for a given galaxy data $f(\theta | {\rm galaxy})$ with respect to the parameter vector $\theta$ is obtained via Bayes' theorem:
\begin{align} \label{bayes}
&f(\theta | {\rm galaxy})
= \frac{f(\theta ) \, \mathcal{L}(\theta )}{\mathcal{E}} \,,\\
&\theta = \left( \loga, \logyb , \logyd, D, I  \right ) \,, \nonumber
\end{align}
where $f(\theta )$ is the prior, $\mathcal{L}(\theta )$ is the likelihood and  $\mathcal{E}$ is the evidence. A brief review on Bayesian inference in this context can be found in the supplementary material of \Rteen{}. We adopt the following set of independent variables for our analysis:
\begin{align}
\loga &\equiv  \log_{10} \left( \frac{a_0}{\rm km/s^2} \right) \, , \\
\logyb &\equiv  \log_{10} \Upsilon_\bulge \, , \label{logybDef}\\ 
\logyd &\equiv  \log_{10} \Upsilon_\disk \, , 	\label{logydDef}
\end{align}
besides galaxy distance $D$ and inclination $I$.

The analysis of this paper focuses on the 1D marginalized posteriors on $\loga$ for each galaxy, which are obtained according to
\begin{equation}
f(\loga| {\rm galaxy})= \int f(\theta  | {\rm galaxy}) \, \d \logyb \, \d \logyd \, \d D \, \d I   \,.
\end{equation}

\subsection{Likelihood}

For each galaxy, we  adopt the Gaussian likelihood
\begin{align}
\mathcal{L}(\theta ) =
|2 \pi \Sigma|^{-1/2}
e^{-\chi^{2}(\theta)/2} \,,
\end{align}
where $\Sigma$ is the covariance matrix, to be detailed below, and $\chi^2 = \chi^2(\loga,\logyb, \logyd, D, I )$, with 
\begin{equation} \label{chi2fun}
\chi^{2}= \sum_{i=1}^{N} \left ( \frac{ V_\model(R_i, \loga, \logyb, \logyd, D) -V_{\Circ, i} \frac{\sin I_0}{\sin I} }{\sigma_{V, i} \frac{\sin I_0}{\sin I} }\right)^2 \, .
\end{equation}
In the above, $N$ is the number of data points of the given galaxy data, $V_\model$ is the model circular velocity, $R_{i}$ is the galaxy radius at which the reference circular velocity $V_{\Circ,i}$ was measured. The corresponding  error is $\sigma_{V,i}$. The quantities $V_{\Circ, i}$ and $\sigma_{V, i}$ are provided by SPARC with the reference inclination $I_0$. The determinant of the (diagonal) covariance matrix is $|\Sigma|= \prod_{i} \sigma_{V,i}^{2}$. 

The model velocity $V_\model$ is the one inferred from MOND using the interpolating function \eqref{int} and the Newtonian velocity \eqref{newtdecompV} with the distance correction \eqref{Dcorrection}.

\subsection{Priors}

As commented in the Introduction, in this work we adopt, for the mass-to-light ratios, distance and galaxy inclinations, essentially the same priors used by \citet{Li:2018tdo} and \citet{Chang:2018lab}. 

Regarding the prior on $\loga$, we adopt a uninformative prior, a flat prior in this case, as MOND does not predict neither a value nor a distribution for $\loga$: it has to be inferred from observational data. In order to facilitate the comparison between our results and the results of other works, we recall that  \citet{Li:2018tdo} considered two different priors for $\loga$, a flat one (just like we are considering here) and a very sharp Gaussian prior centered on the RAR $a_\dagger$  value (which, as expected, strongly restricted any variation on $a_0$ between galaxies). \citet{Chang:2018lab} considered a Gaussian prior on $\loga$, but  they considered it with a much larger dispersion (two orders of magnitude). Their Gaussian width is sufficiently large to include, at $1\sigma$ level, all (or almost all) the $\loga$ posterior modes found in \Rteen{} with a flat prior, hence that Gaussian prior is roughly equivalent to a flat prior.

Decomposing the prior $f(\theta)$ as follows
\begin{equation}
	f(\theta ) = 
f(\loga)f(\logyb)f(\logyd)f(D ) f( I)\, , 
\end{equation}
the priors for each parameter read:
\begin{align}
f(\loga)& = 1/15 \, , \hspace{1.6cm} \mbox{ with } -20 \le \loga \le  -5  \, ,   \label{pr1}\\ 
f(\logyb) &= \mathcal{N}(0.5,0.1^2) \, , \\ 
f(\logyd) &= \mathcal{N}(0.7,0.1^2)  \, , \\ 
f(D )& = \mathcal{N}(D_0,\sigma_D^2)\, , \hspace{0.7cm} \mbox{ with }   D\ge 0.5 \text{ Mpc}   \label{priorD}\, , \\
f(I) &= \mathcal{N}(I_0,\sigma_I^2)\, ,  \hspace{0.9cm} \mbox{ with }     1^\circ \le I \le 90^\circ  \label{priorI}\, .
\end{align}
The errors $\sigma_D$ and $\sigma_I$ vary from galaxy to galaxy according to the values provided by the SPARC dataset. The priors (\ref{pr1}, \ref{priorD}, \ref{priorI}) have  zero value outside the given ranges.  The symbol $\mathcal{N}(x,\sigma_x^2)$ stands for a normal (Gaussian) distribution centered at $x$ and with $\sigma_x$ as the standard deviation (Gaussian root-mean-squared width). In the following, we detail further the priors above:
\begin{enumerate}
	\item[{\bf i) $f(\loga)$.}]  We use a uniform prior with support in a large range. None of the galaxies have 5$\sigma$ posterior distributions with $\loga > -5$, while some galaxies have $\loga < -20$, but such small values are dynamically equivalent to $a_0=0$ (or, $\loga \rightarrow - \infty$); hence extending towards lower and finite values of $\loga$ would be inconsequential. For comparison, the global best value for $\loga$ is about $-13$, while $-20$ is indistinguishable from Newtonian gravity.  
	\item[{\bf ii) $f(\logyb)\mbox{ and } f(\logyd)$.}] For the mass-to-light ratios (which are in the 3.6 $\mu$m band), we use the same central values and dispersions of \citet{McGaugh:2016leg} and \citet{Li:2018tdo} (see also \citet{Meidt:2014mqa} and \cite{2015ApJS..219....5Q}). These priors imply that $\Upsilon_\bulge >0$ and $\Upsilon_\disk >0$, see eqs.~(\ref{logybDef}-\ref{logydDef}).
	\item[{\bf iii) $f(D)$.}]  It is the same prior used by \citet{Li:2018tdo}, but with a constraint. The purpose of the latter is to restrict unrealistic degeneracies with the distance. The closest RAR galaxy has a distance of 0.98 $\pm$ 0.05 Mpc, but some galaxies at $\sim$40 Mpc have 1$\sigma$ uncertainties of $\sim \pm 10$ Mpc. The distances of such galaxies were estimated from the Hubble flow, and this is why they have large relative distance errors.	 
	Eq.~\eqref{priorD} states that none of the RAR galaxies is allowed to be closer than 0.5 Mpc (this is very conservative, Andromeda for instance is at $0.8$ Mpc).
	\item[{\bf iv) $f(I)$.}]  Since the observed inclination is defined in the range from $0^\circ$ to $90^\circ$, being $0^\circ$ a face-on galaxy,  we use the same constraints, but starting from $1^\circ$. We recall that $0^\circ$ corresponds to a singularity (face-on galaxies have no rotation curve), and also that the RAR sample only includes galaxies with $I_0 \ge 30^\circ$. 
\end{enumerate}

The constraints in the priors $f(D)$ and $f(I)$ have no impact on the best-fits, but they can reduce, for a few galaxies, the tails of the $\loga$ posteriors. These constraints are irrelevant for most of the galaxies, they are small deviations from pure Gaussian distributions, and, for few cases, they imply minor improvements from the physical perspective, due to the removal of unrealistic cases. The numerical analysis benefits from these constraints, since they impose a small but finite parameter distance from the singularities $D=0$ and $i=0^\circ$.

\subsection{Quality cuts} \label{quality}

All the galaxies that we consider are from the SPARC sample (175 galaxies). More specifically, we only use those galaxies used to determine the RAR (153 galaxies). Reducing the sample from the SPARC to the RAR sample is the first quality cut that we use ($Q_1$): it eliminates galaxies whose reference inclination value is smaller than $30^\circ$ (i.e., $I_0 < 30^\circ$) and those galaxies with poor concordance between approaching and receding rotation curves, classified as ${\cal Q} = 3$ by \citet{2016AJ....152..157L}. This is also the same quality cut adopted by \citet{Li:2018tdo}. All the data that we provide here is subjected to this quality cut.

The $Q_1$ quality cut uses criteria based on physical data to eliminate galaxies that have a higher chance of providing less accurate $\loga$ determination. We also consider a second quality cut $Q_2$, whose main purpose is the same of $Q_1$, but whose criteria are based on statistical data that suggest that $\loga$ may have not been accurately determined for that galaxy.
$Q_2$ is divided in the following two parts ($Q_2=  Q_{2a} \wedge Q_{2b}$):
\begin{itemize}

\item $Q_{2a}$, the compatibility of the model best fit with the observational data;

\item  $Q_{2b}$, the existence of well-defined 5$\sigma$ credible regions in the range stated in eq.~\eqref{pr1}.

\end{itemize}

Regarding $Q_{2a}$, galaxies with high $\chi^{2}_{\rm min}$, whose $p$-value, considering a $\chi^2$-statistics, is outside the expected 5$\sigma$ region, are eliminated:
\begin{equation}
1 - p\text{-value} \equiv F_{k}(\chi^{2}_{\rm min}) \, \ge \, \erf \frac{n=5}{\sqrt{2}} \,, 
\end{equation}
where $F_{k}(\chi^{2})$ is the cumulative $\chi^2$-distribution with $k$ degrees of freedom, $\chi^{2}_{\rm min}$ is the observed value and $\erf$ is the error function.
In this case it is $k=N-M$ where $N$ is the number of data points and $M$ the number of fitted parameters, which is 4 or 5 depending on the presence of the bulge.%
\footnote{There are 4 galaxies (D512-2, NGC6789, UGC00634, UGC07232) that have $N=M=4$ so that $F_{k}(\chi^{2})$ is singular. If the model was linear, these galaxies would have $\chi^{2}_{\rm min}=0$. As it is well known \citep{Andrae:2010gh}, this does not happen as there are nonlinearities. In order to account for this we consider the effective number of parameters $M_{\rm eff}=M-1$. The consequence of using $M_{\rm eff}$ is that these 4 galaxies pass $Q_{2a}$.}
The reference $p$-value above is $5.7\times 10^{-7}$.
We stress that we only eliminate those with especially high $\chi^{2}_{\rm min}$ values; we do not assume that $\chi^{2}_{\rm min}$ values provide a good standard for model comparison or quality of fits in general \citep{Andrae:2010gh}; the assumption is that a too high $\chi^2$ value is a qualitative sign that possibly something is wrong; if the model is assumed right, then the data is possibly problematic, therefore eliminating it may be safer.

For $Q_{2b}$, there are some galaxies for which $a_0=0$ is compatible with the data at 5$\sigma$ level, leading to credible regions on $\loga$ that are not bounded from below. These galaxies would not improve the chances of a fundamental $a_0$ value, being ``outliers'' as far as MOND is concerned.
$Q_{2a}$ and $Q_{2b}$ are evaluated for each galaxy and the results are shown in Table \ref{tab:individual}.

Although we always use $Q_1$, we evaluate our results both with and without $Q_2$. Therefore, $Q_2$ provides additional support that our results are robust against outliers, while being clear that our results do not depend on $Q_2$.

At last, we also stress that these quality cuts are applied homogeneously to all the sample. That is, we never re-analyze {\it a posteriori}  the observational data, looking for case by case justifications for why particular galaxies did not provide some expected result. Any sample may always be subjected to some unknown relevant systematics, but our hope is that the observational data is being properly handled, with at most a few ill-modeled cases that would not contaminate the complete statistics. The $Q_2$ quality cut helps on providing an additional safety measurement against unknown systematics, and its application here implies a decrease on the sample size from 153 galaxies (the RAR sample) to 91 galaxies. This quality cut was introduced, in this galaxy context, in \Rteen{}.

The sample composed by the SPARC galaxies with the $Q_1$ quality cut is labeled ${\cal S}_1$ (the RAR sample), while the sample with the two quality cuts, $Q_1 \wedge Q_2$, is labeled ${\cal S}_2$.

\subsection{Global best value} \label{gbv}

After applying some set of quality cuts, one is left with $N_{\rm G}$ galaxies, indexed by $g$, from a subset ${\cal S}$ of the SPARC sample. Assuming a common $\loga$ value, we combine all the galaxy as follows:
\begin{align}
&f(\loga,\theta_1, ..., \theta_{N_{\rm G}} | {\cal S})
= \frac{f(\loga, \{ \loga_g\}) \prod_{g=1}^{N_{\rm G}} f(\theta_g)   \mathcal{L}(\theta_g )}{\mathcal{E}} \,, \\
&f(\loga, \{ \loga_g\}) = \delta (\loga-\loga_1) \dots \delta (\loga-\loga_{N_{\rm G}}) \,, \label{priorFundamental} \\
&\theta_g = ( \loga_g,  \logyb_g, \logyd_g, D_g, I_g  ) \,. \nonumber
\end{align}
where we used the fact that the galaxies' rotation curves are independent from each other.
The vector $\theta_g$ is the analogue of $\theta$, but indexed for each one of the galaxies.  Priors and likelihoods are as in Eq.~\eqref{bayes}.

The prior of Eq.~\eqref{priorFundamental} imposes that all the $\loga_g$ are equal to $\loga$, and thus among themselves.
One can then marginalize over the $N_{\rm G}$ variables $\loga_{g}$ so that one obtains a posterior that depends on $\loga$. Indeed, introducing
\begin{equation}
	\hat \theta_g = ( \logyb_g, \logyd_g ,D_g, I_g  )\, ,
\end{equation}
we write,
\begin{align}
f(\loga,  \hat \theta_1 ... \hat \theta_{N_{\rm G}} | {\cal S})
&= \int \d \loga_1 ... \d \loga_{N_{\rm G}} \, f(\loga,\theta_1, ..., \theta_{N_{\rm G}} | {\cal S})   \nonumber \\
&=\frac{f(\loga )^{N_{\rm G}}}{\mathcal{E}}  \prod_g  f(\hat \theta_g )   \mathcal{L}(\loga, \hat \theta_g ) \,.
\end{align}
Finally, it follows that the marginalized posterior on $\loga$ is  proportional to the product of the marginalized posteriors on $\loga$ from the individual galaxies:
\begin{align}
&f(\loga| {\cal S}) = \frac{f(\loga )^{N_{\rm G}}}{\mathcal{E}}  \prod_g \int \d \hat \theta_g  f(\hat \theta_g )   \mathcal{L}(\loga, \hat \theta_g ) \nonumber \\
&= \frac{\prod_g \mathcal{E}_g}{\mathcal{E}}  \prod_g f(\loga| {\rm galaxy}_{\!g})  
\equiv \frac{C}{\mathcal{E}} \prod_g f(\loga| {\rm galaxy}_{\!g})  \,, \label{mara}
\end{align}
where the evidences $\mathcal{E}_g$ are defined in eq.~\eqref{bayes}, and we defined the  constant $C$.
We call  ``global best value'', and denote it by $\loga^{\rm gbv}$, the value of $\loga$ that maximizes $f(\loga| {\cal S})$. That is, it is the mode of the $\loga$  distribution marginalized over all the other parameters.%
\footnote{See \citet{deAlmeida:2018kwq} for a similar method, albeit for a different model.}
A comment supporting this way of finding the best $a_0$ can also be found in \citet{1988ApJ...333..689M}.
Note that this method (see the prior \eqref{priorFundamental}) assumes rather than tests the existence of a universal parameter.
In the following, we will instead assess the compatibility between the various galaxy posteriors on $\loga$.

In \Rteen{} we computed the global best value by approximating the posteriors $f(\loga| {\rm galaxy}_{\!g})$ as Gaussians and performing a $\chi^2$ minimization. In practice, for this application, the difference is very small (see Table \ref{tab:res}). However, we stress that the method here used is to be preferred to that of \Rteen{}, since it does not rely on a Gaussian approximation for the mean and variance. Here we use the full posterior distribution up to 5$\sigma$. However, to fully compare the posteriors in the full range $-20 \le \loga \le -5$, we need to extend the posteriors beyond the $5\sigma$ level, and a Gaussian extension is used to this end. This extension is detailed in Appendix \ref{app:ext}. In Appendix \ref{app:revGau} we briefly review the Gaussian approximation that was used in \Rteen{}.

\subsection{$X^2$ statistics} \label{X2stat}

We will now discuss how to quantify the compatibility between the posteriors $f(\loga| {\rm galaxy}_{\!g})$.
In \citetalias{Rodrigues:2018duc} we adopted the $\chi^2$ statistic of equation~\eqref{chi2a0}, whose $\chi^{2}_{\rm min}$ follows a $\chi^{2}$ distribution with $k=N_g-1$ degrees of freedom. Here, we  improve upon the method of \citetalias{Rodrigues:2018duc}, performing a more accurate procedure that does consider the full posteriors $f(\loga| {\rm galaxy}_{\!g})$, instead of using Gaussian approximations based on the mean and the variance of $f(\loga| {\rm galaxy}_{\!g})$.
Our new method is based on the Bayes factor (the ratio between two Bayesian evidences) and extends the $\chi^2$ statistics to the general non-Gaussian case, reducing to it when the distributions $f(\loga|{\rm galaxy}_{\!g})$ are Gaussian.
We name this generalization the $X^2$ statistics. It was inspired by the results of \citet{Verde:2013wza,Lin:2017ikq}.

First, following the notation of \citet{Verde:2013wza,Lin:2017ikq}, we introduce the  Tension estimator $\mathcal{T}$, which  is the Bayes factor defined as the ratio between a virtual hypothesis (Bayesian evidence $\bar{\mathcal{E}}$, see below) and the actual hypothesis (evidence $\mathcal{E}$):

\begin{figure*}
\centering
\includegraphics[height=7.5cm ]{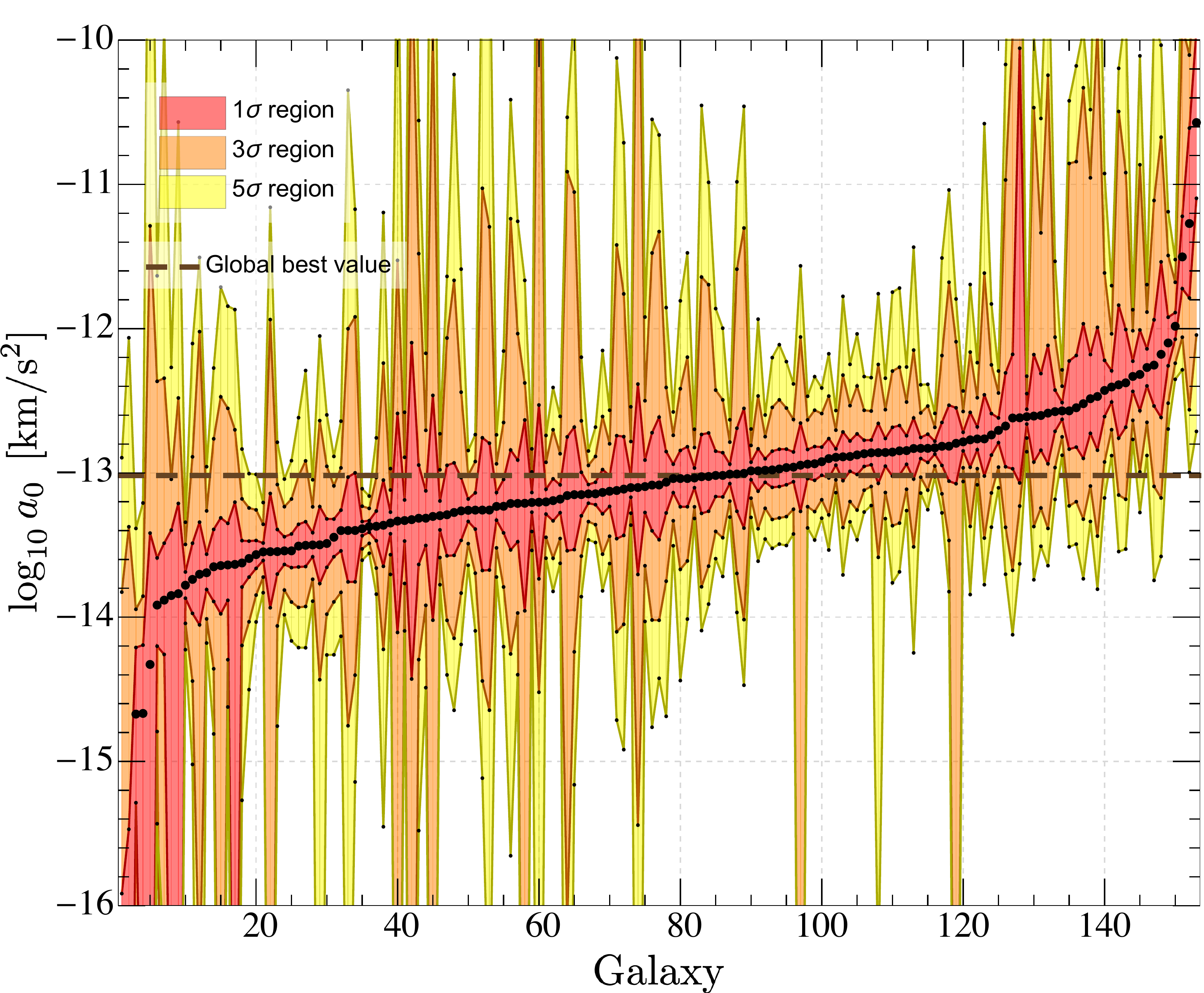} \hspace*{-0.26cm}
\includegraphics[trim={3.0cm 0 0 0},clip, height=7.5cm ]{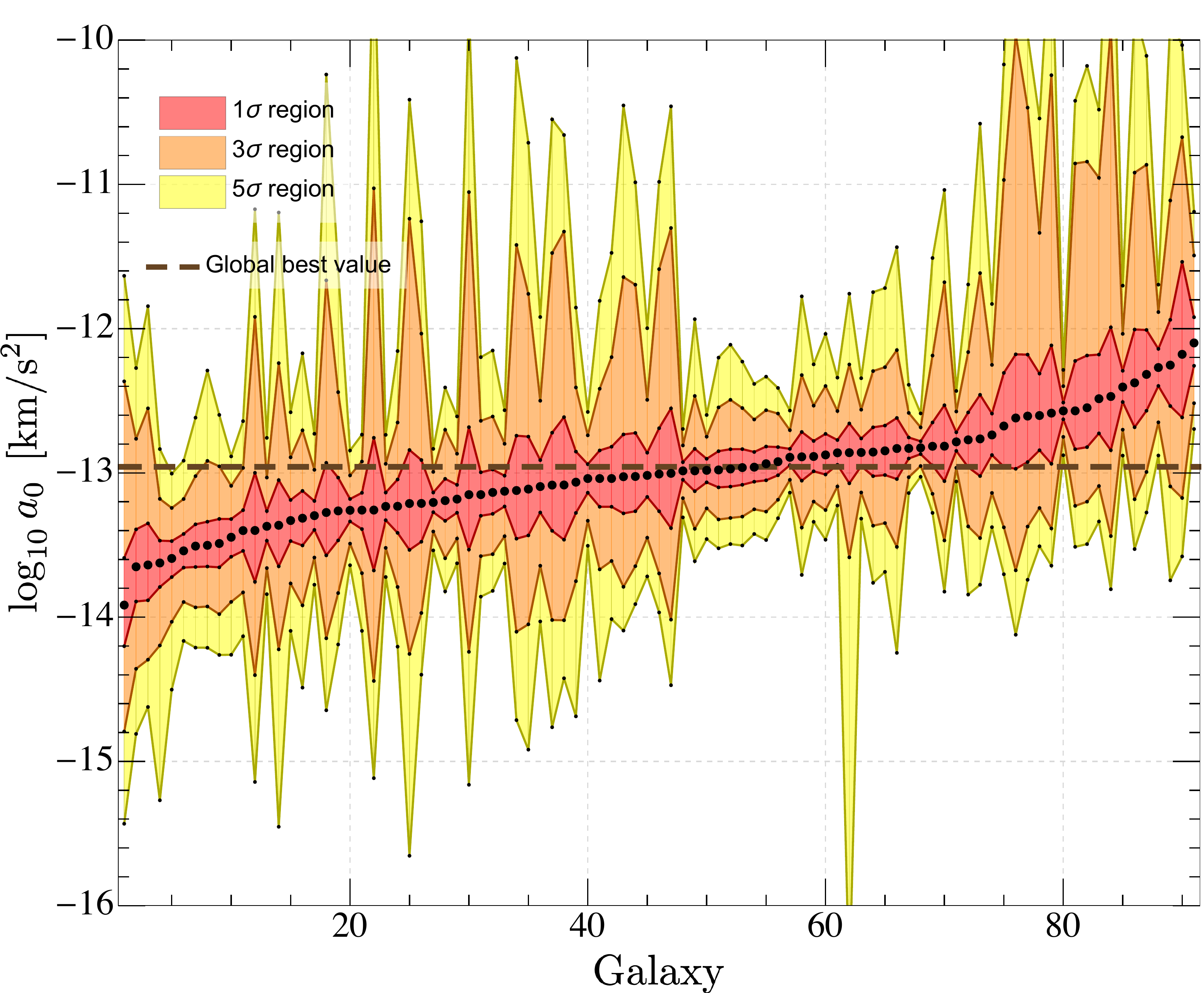}
\caption{ Posterior probability distributions of $\loga \equiv \log_{10} [a_0 /({\rm km/s^2})]$ for each galaxy of the SPARC database that passed the quality cuts $Q_1$ (left, with 153 galaxies, the RAR sample ${\cal S}_1$) and $Q_1 \wedge Q_2$ (right, with 91 galaxies, the sample ${\cal S}_2$). Each posterior, for each galaxy, has been obtained after marginalizing over stellar mass-to-light ratios, galaxy distance and galaxy inclination, and are displayed showing the maximum (mode) of the posterior (large black dots) together with the 1, 3, and 5$\sigma$ credible intervals (red, orange and yellow shaded regions, respectively). To enhance the clarity of the plot, the galaxies are sorted according to their posterior maximum, and small black dots are used to display the limits of the credible intervals for each galaxy. The global best value of $\loga$, for each of the cases, is shown with a dashed line: it is evident that many galaxies are not compatible with its value.
}
\label{fig:bands}
\end{figure*}

%
\begin{align} \label{tension}
\mathcal{T} &= \frac{\bar{\mathcal{E}}}{\mathcal{E}} \,.
\end{align}
We remind the reader that the Bayes factor gives the odds ratio relative to the two hypotheses and is a standard tool of Bayesian inference \citep[see, e.g.,][]{gregory2010bayesian}.

The evidence $\mathcal{E}$ is proportional to the integral of the product of the posteriors, that is,
\begin{align} \label{E1}
\mathcal{E} = C \int \d \loga \prod_g  f(\loga| {\rm galaxy}_{\!g}) \, ,
\end{align}
where $C$ is the  constant of eq.~\eqref{mara}. The value of $C$ will not be relevant, as it will become clear shortly.

Similarly, the evidence $\bar{\mathcal{E}}$ is  obtained according to
\begin{align}
\bar{\mathcal{E}} = C \int \d \loga \prod_g  \bar f(\loga| {\rm galaxy}_{\!g}) \,,
\end{align}
where $\bar f(\loga| {\rm galaxy}_{\!g})$ is the posterior of the $g$-th galaxy translated by the difference $\loga^{\rm gbv}- \loga_{g}^\text{mean}$ to a common reference point, the global best value $\loga^{\rm gbv}$:
\begin{align}
\bar f(\loga| {\rm galaxy}_{\!g}) =  f(\loga+\loga_{g}^\text{mean} - \loga^{\rm gbv}  | {\rm galaxy}_{\!g}) \,.\label{transla}
\end{align}
Note that $C$ cancels out in equation~\eqref{tension}. 
Eq.~\eqref{transla} implies that the mean $\loga_g$ according to the distribution $\bar f$ coincides with $\loga^{\rm gbv}$. Therefore, the translated posteriors $\bar f(\loga| {\rm galaxy}_{\!g})$ overlap among themselves more than the original ones so that  $\bar{\mathcal{E}} \ge \mathcal{E}$.
The idea behind the Tension is to build an estimator that is sensitive to the degree of overlapping between different posteriors.
In order to maximize the sensitivity of the Tension, one should translate the distributions in order to have $\mathcal{T}\ge 1$, that is, maximize the overlapping and so $\bar{\mathcal{E}}$.
This is approximately achieved by translating the posteriors according to their distribution means $\loga_{g}^\text{mean}$ as in eq.~\eqref{transla}. 
One could consider using the mode of $\loga_g$, instead of the mean value; however, with the purpose of using Monte Carlo to simulate the expected distribution, it is preferable to use the mean value so that the MC data have a mean that reproduces the desired value. Nonetheless, we have verified  that the difference between these choices is negligible, at least for this application.

The statistic $X^2$ is defined from the identification
\begin{align} \label{Xdef}
X^2 \equiv 2 \ln \mathcal{T}\ge0 \,,
\end{align}
which reduces to the $\chi^2$ statistic of eq.~\eqref{chi2a0} if the distributions are Gaussian \citep{Lin:2017ikq}:
\begin{align}
X^2 \longrightarrow \chi^{2}_{\rm min} \,, 
\end{align}
so that $X^2$ follows a $\chi^{2}$ distribution with $k=N_{\rm G}-1$ degrees of freedom.
However, $X^2$ is well defined also in the general non-Gaussian case as one can compute the evidences without any approximation.

Now, in order to assess the significance of a value of~$X^2$, we need the distribution of the $X^2$ statistic, which is expected to resemble the $\chi^{2}$ distribution.
In the general case it is not possible to obtain it analytically and one has to resort to a Monte Carlo simulation.
The $X^2$ distribution is obtained under the null hypothesis that a fundamental $\loga$ exists.
To this end we generate for each galaxy~$g$ a random value $\loga_g^{\rm null}$ drawn from the numerical distribution $\bar f(\loga| {\rm galaxy}_{\!g})$: we use the full non-Gaussian distribution whose mean is at $\loga^{\rm gbv}$ in agreement with the null hypothesis.
We then translate the distribution so that its mean is $\loga_g^{\rm null}$:
\begin{align}
f_{\rm null}(\loga| {\rm galaxy}_{\!g}) =  f(\loga+\loga_g^{\rm mean} - \loga_g^{\rm null}  | {\rm galaxy}_{\!g}) \,.
\end{align}

These posteriors can then be used to compute
\begin{align} \label{tension2}
&\mathcal{E}^{\rm null} = C \int \d \loga \prod_g  f_{\rm null}(\loga| {\rm galaxy}_{\!g}) \,, \\
&X^2_{\rm null} =2 \ln \frac{\bar{\mathcal{E}}}{\mathcal{E}^{\rm null} }\,.
\end{align}
By repeating many times the above algorithm one can generate numerically the $X^2$ distribution, from which one can compute the confidence in rejecting the null hypothesis.
Although expensive, this method does not introduce any approximation and reduces to the well-known $\chi^{2}$ distribution when the non-Gaussianity of the posteriors is small.

\section{Numerical methods}\label{NumMethods}

Our numerical procedure consists of the following pipeline divided into three stages:
\begin{enumerate}
	\item  The first stage loads the SPARC data, sets the likelihood functions and priors for each galaxy \eqref{bayes},  runs  optimization procedures  to find the best fit for each galaxy, exports the results. This stage is done with the \texttt{MAGMA} package.\footnote{\href{https://github.com/davi-rodrigues/MAGMA}{github.com/davi-rodrigues/MAGMA} \citep{Rodrigues:2018duc}}

	\item The second stage is the MCMC itself. After importing the last stage data, we  use \texttt{emcee},\footnote{\href{https://github.com/dfm/emcee}{github.com/dfm/emcee} \citep{ForemanMackey:2012ig}} an affine invariant sampler for Markov chain Monte Carlo (MCMC), to sample the posteriors $f(\theta| {\rm galaxy}_{\!g})$. The best fits from the previous stage are used to set the starting conditions of the MCMC, reducing the needed computational time (we adopt a burn-in of 10\%).
Specifically, we computed chains of  $50 \times 10^{6}$ total points from 100 walkers, which is more than 100 times the autocorrelation time. The generated chains are exported.
We have  crosschecked that $45 \times 10^{6}$ points are enough in order to obtain reliable $5\sigma$ credible intervals (see Appendix~\ref{app:mcmc})
 by comparing the $\loga$ posteriors from \texttt{emcee} with those obtained via grid evaluation using \texttt{mBayes},\footnote{\href{https://github.com/valerio-marra/mBayes}{github.com/valerio-marra/mBayes} \citep{Camarena:2018nbr}} which computes arbitrarily high credible intervals with negligible error.

	\item The third stage analyzes the chains. We use \texttt{getdist}%
\footnote{\href{https://github.com/cmbant/getdist}{github.com/cmbant/getdist} \citep{Lewis:2019xzd}} for obtaining the credible intervals and the triangular plots (see, e.g., Figure~\ref{fig:triplot}). 
The analysis of the $X^2$ statistics is carried out using specialized numerical code for the two samples ${\cal S}_1$ and ${\cal S}_2$ discussed in Section~\ref{quality}.
The $X^2$ distribution is obtained with a Monte Carlo (MC) simulation with $62 \times 10^6$ MC points for ${\cal S}_1$ and $123 \times 10^6$ MC points for  ${\cal S}_2$.

\end{enumerate}

\section{Results}\label{results}

\begin{table*}
\caption{Results for individual galaxies. The full table is provided as machine readable supplementary material. The columns are: (1) Galaxy name. (2) Best fit for $A_0 = \log_{10} a_0 \; [{\rm km/s}^2]$. (3) Minimum $\chi^2$ value. (4) Number of rotation curve data points. (5) Part $a$ of the quality cut $Q_2$. (6) Part $b$ of the quality cut $Q_2$.
A `1' is used if the galaxy passes a quality cut and 0 is used otherwise. In order to pass the $Q_2$ quality cut, both ${Q}_{2a}$ and ${Q}_{2b}$ need to be 1. 
(7) The mode of the marginalized $A_0$ posterior. (8-13) The limits of the $A_0$ credible intervals, with respect to the $A_0$ mode, for 1$\sigma$, 3$\sigma$ and 5$\sigma$.}
\centering
\renewcommand{\arraystretch}{1.5}
\begin{tabular}{lrrrrrrrrrrrr}
\hline\hline
Galaxy     &$A_0$ best fit &$\chi^2_{\rm min}$    &$N$ &${Q}_{2a}$ &$Q_{2b}$ &$A_0$ mode &1$\sigma_-$   &1$\sigma_+$  &3$\sigma_-$   &3$\sigma_+$  &5$\sigma_-$   &5$\sigma_+$  \\[-0.15cm]
(1)&(2)&(3)&(4)&(5)&(6)&(7)&(8)&(9)&(10)&(11)&(12)&(13)\\
\hline
CamB       &-14.645 &38.884  &9  &1  &0   &-14.672 &-0.614&0.462&-5.328&0.727&-5.328&1.287\\
D512-2     &-13.241 &0.297   &4  &1  &0   &-13.258 &-0.416&0.463&-1.388&1.963&-3.752&4.179\\
D564-8     &-13.486 &8.546   &6  &1  &1   &-13.502 &-0.147&0.164&-0.424&0.585&-0.710&1.212\\
D631-7     &-13.140 &183.494 &16 &0  &1   &-13.145 &-0.075&0.080&-0.214&0.259&-0.337&0.462\\
DDO064     &-13.030 &3.712   &14 &1  &1   &-13.085 &-0.317&0.363&-0.935&1.609&-1.679&2.535\\
...	     &... &...   &... &...  &...   &... &...&...&...&...&...&...\\
\hline \hline
\end{tabular}
\label{tab:individual}
\end{table*}

\begin{figure}
\centering
\includegraphics[width= \columnwidth]{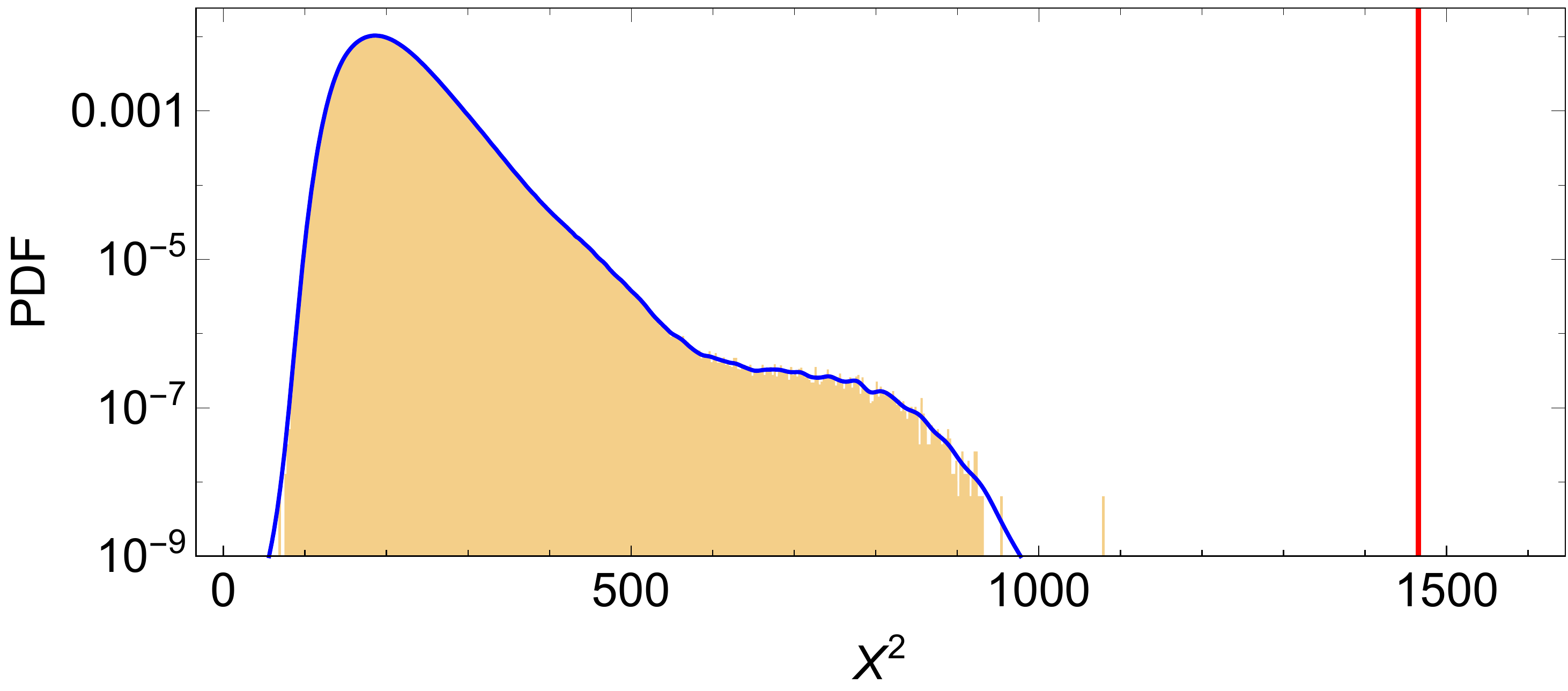}
\includegraphics[width= \columnwidth]{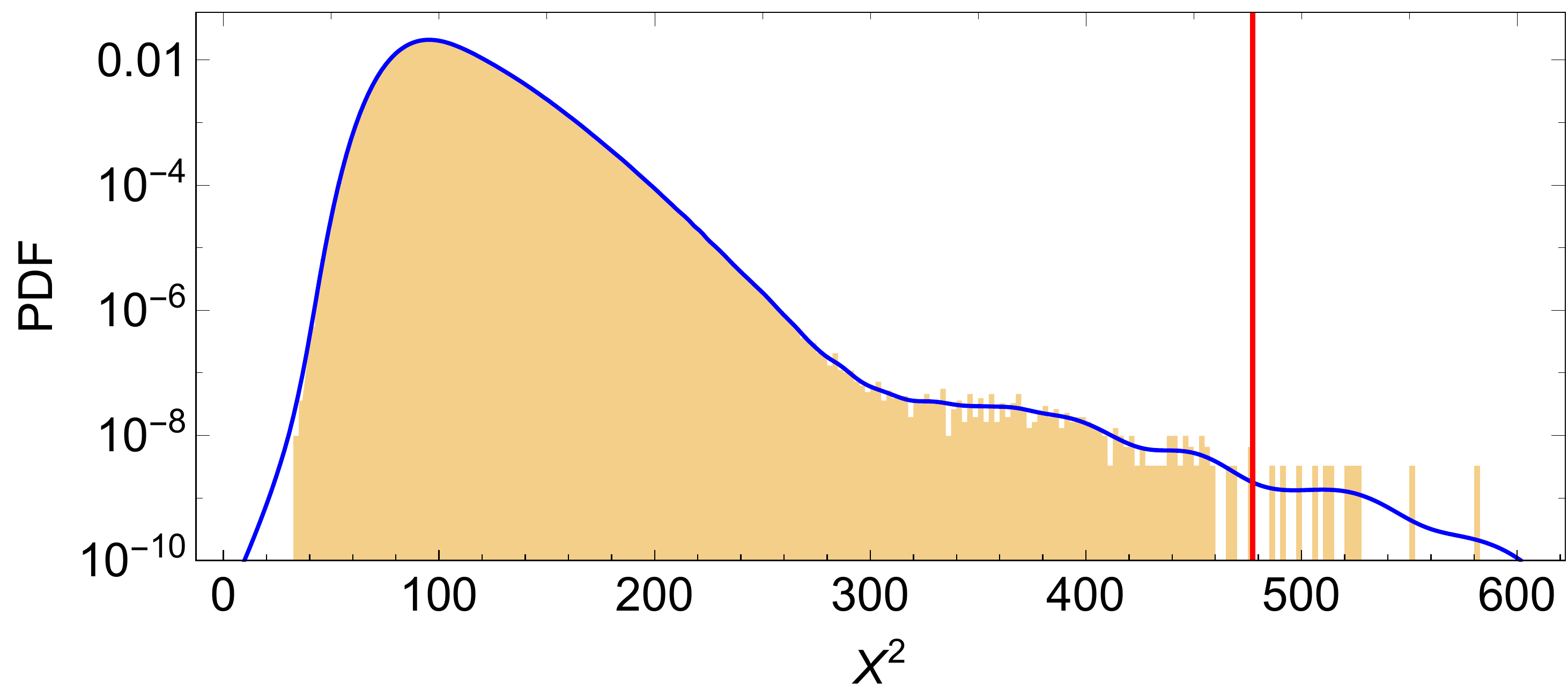}
\caption{Numerical distribution of the $X^2$ statistic that generalizes the $\chi^2$ statistic to the case of non-Gaussian distributions.
The filled yellow histogram was obtained numerically using a Monte Carlo simulation. The blue solid line is the smoothed empirical distribution. The vertical red line marks the values $X^2$ reported in Table~\ref{tab:res}. The hypothesis that there is a fundamental acceleration scale is ruled out with the confidence of $>5.7\sigma$ for ${\cal S}_1$ and of~$5.3\sigma$ for ${\cal S}_2$.}
\label{fig:tension}
\end{figure}

We will now present our results for the two quality cuts discussed in Section~\ref{quality}.
Figure~\ref{fig:bands} shows modes and 1, 3 and 5$\sigma$ credible intervals of the posteriors $f(\loga| {\rm galaxy}_{\!g})$ for the 153 galaxies that passed the quality cut $Q_1$ and the 91 galaxies that passed the combined  $Q_1 \wedge Q_2$ quality cut: no single value of $a_0$ cuts through all the $5\sigma$ credible intervals.
Also shown with a dashed line is the global best value of Section~\ref{gbv}.
From Figure~\ref{fig:bands} it is clear that many galaxies are not quite compatible with the global best value.
The numerical values are given in Table~\ref{tab:res}.

\begin{table}
\caption{Numerical results for the two samples ${\cal S}_1$ and ${\cal S}_2$. Top: results with the tension-based method as here proposed. Bottom: for comparison's sake we report also the results that adopt the Gaussian approximation, following the \Rteen{} methodology.
The hypothesis that there is a fundamental acceleration scale is ruled out with a confidence larger than $5\sigma$.
}
\centering
\renewcommand{\arraystretch}{1.5}
\begin{tabular}{lcc}
\hline\hline
Tension-based method &  ${\cal S}_1$ (RAR sample) &  ${\cal S}_2$ \\
\hline 
$a_0^{\rm gbv}$ [km/s$^2$] & $0.96 \times 10^{-13}$ & $1.10 \times 10^{-13}$  \\
$X^2$ & 1465 & 477 \\
rejection of fundamental $a_0$ & $>5.7\sigma$ & $5.3\sigma$ \\
\hline\hline
\Rteen{} Gaussian method & ${\cal S}_1$ (RAR sample) & ${\cal S}_2$ \\
\hline 
$a_0^{\rm gbv} $ [km/s$^2$] & $0.91 \times 10^{-13}$ & $1.06 \times 10^{-13}$ \\
$\chi^{2}_{\rm min}$ & 1280 & 438 \\
rejection of fundamental $a_0$ & $28 \sigma$ & $14 \sigma$ \\
\hline\hline
\end{tabular}
\label{tab:res}
\end{table}%

Also in Table~\ref{tab:res} we report the values of the $X^2$ statistic.
For the sake of comparison we also report the values in the Gaussian approximation, under which $X^2 \rightarrow \chi^{2}_{\rm min}$.
 Under this approximation one can compute the confidence of rejecting the hypothesis that there is a fundamental acceleration scale using the $\chi^2$ distribution. For the sample ${\cal S}_2$ one finds $14\sigma$.
If we approximate the $\chi^2$ distribution itself as a Gaussian, as done in \citetalias{Rodrigues:2018duc}, the confidence is 26$\sigma$. This is due to the fact that the tail of the $\chi^2$ distribution is longer than the one of the corresponding Gaussian. 

About the results above using the Gaussian approximation, we comment that this latter value (26$\sigma$) is in qualitative agreement with the results of \citet{Chang:2018lab}. They considered the quality cut $Q_2$, but not in the same way we are doing here. They eliminated the same galaxies that were eliminated in \Rteen{} due to that quality cut $Q_2$ in that reference (leaving a sample size of 100 galaxies). However, \Rteen{} uses different priors, and the $Q_2$ quality cut depends on the priors. Here we reselect the galaxies to be removed considering the quality cuts, thus the resulting sample of galaxies, when $Q_2$ is applied, is different with respect to \Rteen{} and \citet{Chang:2018lab}. This difference on the sample of galaxies  naturally leads to a difference on the face-value confidence in rejecting the fundamental acceleration scale (although, we stress, there is no conflict on the conclusions). At last, we also recall that the priors used by \citet{Chang:2018lab} are very similar to the priors of \citet{Li:2018tdo} and the ones that we use here, but they are not identical, which may be the reason of some differences on the length of the 5$\sigma$ credible intervals.%
\footnote{The  $5\sigma$ credible intervals that we find here are on average slightly larger than those of \citet{Chang:2018lab}.
Besides the use of slightly different priors, the discrepancy could be due to the fact that one needs a large number of MCMC points in order to reliably sample 5$\sigma$ credible intervals.}

Here we go beyond the Gaussian approximation of \citetalias{Rodrigues:2018duc} and therefore  we adopt the  non-Gaussian $X^2$ value.
In order to assess its significance, we compute the $X^2$ distribution according to the null hypothesis that a fundamental acceleration scale exists. We adopt the Monte Carlo simulation described in Sections~\ref{X2stat} and \ref{NumMethods}.
The result is given in Figure~\ref{fig:tension}: the observed $X^2$ values are shown with a red line. It is evident that they lie deep into the tail of the distribution, which is  modeled in a robust way.
Finally, we estimate the PDF by smoothing the empirical distribution (in light orange) via an adaptive Gaussian kernel. The result is shown with a blue line in Figure~\ref{fig:tension}.

Regarding ${\cal S}_1$, none of the MC points fall near the observed value. As we generated $62 \times 10^6$ points and none was greater than the observed one, we can conclude, very conservatively, that the null hypothesis is ruled out at~$>5.7\sigma$.
Regarding ${\cal S}_2$, out of the total $123 \times 10^6$ MC points, we obtained 11 MC values greater than the observed one so that we can conclude that the null hypothesis is ruled out with the high confidence of~$5.3\sigma$.
Using the smoothed distribution (blue curve in Figure~\ref{fig:tension}) we obtain the same significance.
The significance is lower as compared to the Gaussian case because the non-Gaussian distributions of individual galaxies features, on average, longer tails \citep{Rodrigues:2020squ, 2020NatAs.tmp...15C}.

\section{Conclusions}\label{conclusions}

This work stresses further the importante of testing the compatibility between the acceleration scales derived from the individual galaxies, denoted by  $\loga_g = \log_{10} a_{0 g } \; [\mbox{km/s}]$, where $g$ is an index that labels the galaxies. Such test is important for understanding the meaning of MOND and the RAR, and it is a fundamental test for MOND, since it tests one of the cornerstone assumptions of MOND (the universality of $a_0$) in the context that most favors   MOND, that of rotationally supported galaxies. This work improves the methodology of \Rteen{}, confirming its results that, in the context of rotationally supported galaxies, the Radial Acceleration Relation (RAR) is an emergent correlation. To achieve this conclusion, in summary,  we assume that the function that expresses the RAR correlation (the interpolating function) is common to all the galaxies, but the value of the RAR acceleration scale $a_\dagger$ is not used. For the acceleration scale we use a free constant $a_{0 g}$ for each galaxy.
Then we test, using Bayesian inference, whether the individual $a_{0 g}$ constraints found for each one of the galaxies are compatible among themselves (this requires the computation of the marginalized posteriors on $a_{0 g}$, and this cannot be found from best fits alone).  With the improved methods presented here, we find that the RAR does not imply a fundamental property of (rotationally supported) galaxies but rather an emergent behavior. In this way, we have confirmed  that there is a strong evidence against a common $a_0$ value:  a fundamental acceleration scale $a_0$   is  rejected at more than~5$\sigma$.

This behaviour is not in contradiction with standard dark matter interpretations, but it is at odds with MOND, if the latter is assumed to be a fundamental law at the level of (rotationally supported) galaxies. If MOND is viewed as an empirical approximation relation, then MOND is essentially the RAR: a phenomenological and simple rule valid for galaxies that captures certain average dynamics whose underlying physics is more complex. From this perspective, it is a useful correlation that, at the moment, it has no clear contradiction with the standard dark matter picture \citep{2019ApJ...882....6S}, but see \citet{Ren:2018jpt} for possible hints on an underlying non-standard dark matter physics. Also, some of the RAR features can be seen as necessary consequences of galaxy evolution in a dark matter context \citep{Ludlow:2016qzh, Navarro:2016bfs, Dutton:2019gor}.

On the new methodology used here, besides using essentially the same priors of \citet{Li:2018tdo}, we go beyond the Gaussian approximation that \Rteen{} adopted in order to quantify the tension between the posteriors on $a_0$. Here we have introduced the $X^2$ statistics, based on the  Tension estimator $\mathcal{T}$ proposed by  \citet{Verde:2013wza}, which in turn is based on the Bayes factor \citep[e.g.,][]{gregory2010bayesian, 9781447494782}.
The $X^2$ statistics extends the $\chi^2$ statistics to the general non-Gaussian case, reducing to it when the variables follow a Gaussian distribution.
We expect the $X^2$ statistics to have useful applications in other contexts.

In order to rule out the possibility that our results are driven by outliers, we have carried out the analysis using two different quality cuts. The first is the same quality cut used to evaluate the RAR (leading to what we call the RAR sample). Using this quality cut we found that the hypothesis that there is a fundamental acceleration scale is ruled out with the a confidence clearly larger than $5.7\sigma$.
Then we considered a stronger quality cut based on statistical arguments that could indicate a lower chance of determining an accurate $\loga$ value. This is a strong quality cut that reduced the sample size by $\sim 60\%$, with many cases of very low and very high $\loga$ values being removed. But even with this quality cut a fundamental acceleration scale is ruled out with a confidence larger than $5\sigma$.
Therefore, our results are robust also regarding outliers.

\section*{Acknowledgements}
This work made use of the CHE cluster, managed and funded by COSMO/CBPF/MCTI, with financial support from FINEP and FAPERJ, and operating at the Javier Magnin Computing Center/CBPF. We thank Martin Makler for providing access and help to the CHE cluster. This work also made use of the Virgo Cluster at Cosmo-ufes/UFES, which is funded by FAPES and administrated by Renan Alves de Oliveira. This work made use of SPARC (Spitzer Photometry \& Accurate Rotation Curves). VM and DCR also thank CNPq (Brazil) and FAPES (Brazil) for partial financial support.


\bibliography{mond} %

\begin{thebibliography}{}
\makeatletter
\relax
\def\mn@urlcharsother{\let\do\@makeother \do\$\do\&\do\#\do\^\do\_\do\%\do\~}
\def\mn@doi{\begingroup\mn@urlcharsother \@ifnextchar [ {\mn@doi@}
  {\mn@doi@[]}}
\def\mn@doi@[#1]#2{\def\@tempa{#1}\ifx\@tempa\@empty \href
  {http://dx.doi.org/#2} {doi:#2}\else \href {http://dx.doi.org/#2} {#1}\fi
  \endgroup}
\def\mn@eprint#1#2{\mn@eprint@#1:#2::\@nil}
\def\mn@eprint@arXiv#1{\href {http://arxiv.org/abs/#1} {{\tt arXiv:#1}}}
\def\mn@eprint@dblp#1{\href {http://dblp.uni-trier.de/rec/bibtex/#1.xml}
  {dblp:#1}}
\def\mn@eprint@#1:#2:#3:#4\@nil{\def\@tempa {#1}\def\@tempb {#2}\def\@tempc
  {#3}\ifx \@tempc \@empty \let \@tempc \@tempb \let \@tempb \@tempa \fi \ifx
  \@tempb \@empty \def\@tempb {arXiv}\fi \@ifundefined
  {mn@eprint@\@tempb}{\@tempb:\@tempc}{\expandafter \expandafter \csname
  mn@eprint@\@tempb\endcsname \expandafter{\@tempc}}}

\bibitem[\protect\citeauthoryear{Andrae, Schulze-Hartung  \& Melchior}{Andrae
  et~al.}{2010}]{Andrae:2010gh}
Andrae R.,  Schulze-Hartung T.,   Melchior P.,  2010,
  [\href{https://arxiv.org/abs/1012.3754}{1012.3754}].

\bibitem[\protect\citeauthoryear{Camarena \& Marra}{Camarena \&
  Marra}{2018}]{Camarena:2018nbr}
Camarena D.,  Marra V.,  2018, \mn@doi [Phys. Rev.]
  {10.1103/PhysRevD.98.023537}, D98, 023537,
  [\href{https://arxiv.org/abs/1805.09900}{1805.09900}].

\bibitem[\protect\citeauthoryear{{Cameron}, {Angus}  \& {Burgess}}{{Cameron}
  et~al.}{2020}]{2020NatAs.tmp...15C}
{Cameron} E.,  {Angus} G.~W.,   {Burgess} J.~M.,  2020, \mn@doi [\natastron]
  {10.1038/s41550-019-0998-2}, p.~15.

\bibitem[\protect\citeauthoryear{Chang \& Zhou}{Chang \&
  Zhou}{2019}]{Chang:2018lab}
Chang Z.,  Zhou Y.,  2019, \mn@doi [\mnras] {10.1093/mnras/stz961}, 486, 1658,
  [\href{https://arxiv.org/abs/1812.05002}{1812.05002}].

\bibitem[\protect\citeauthoryear{Dutton, Macci{\`o}, Obreja  \& Buck}{Dutton
  et~al.}{2019}]{Dutton:2019gor}
Dutton A.~A.,  Macci{\`o} A.~V.,  Obreja A.,   Buck T.,  2019, \mn@doi [\mnras]
  {10.1093/mnras/stz531}, 485, 1886,
  [\href{https://arxiv.org/abs/1902.06751}{1902.06751}].

\bibitem[\protect\citeauthoryear{Famaey \& McGaugh}{Famaey \&
  McGaugh}{2012}]{Famaey:2011kh}
Famaey B.,  McGaugh S.,  2012, \mn@doi [Living Rev. Rel.]
  {10.12942/lrr-2012-10}, 15, 10,
  [\href{https://arxiv.org/abs/1112.3960}{1112.3960}].

\bibitem[\protect\citeauthoryear{{Fattahi}, {Navarro}, {Frenk}, {Oman},
  {Sawala}  \& {Schaller}}{{Fattahi} et~al.}{2018}]{2018MNRAS.476.3816F}
{Fattahi} A.,  {Navarro} J.~F.,  {Frenk} C.~S.,  {Oman} K.~A.,  {Sawala} T.,
  {Schaller} M.,  2018, \mn@doi [\mnras] {10.1093/mnras/sty408}, 476, 3816,
  [\href{https://arxiv.org/abs/1707.03898}{1707.03898}].

\bibitem[\protect\citeauthoryear{Foreman-Mackey, Hogg, Lang  \&
  Goodman}{Foreman-Mackey et~al.}{2013}]{ForemanMackey:2012ig}
Foreman-Mackey D.,  Hogg D.~W.,  Lang D.,   Goodman J.,  2013, \mn@doi [Publ.
  Astron. Soc. Pac.] {10.1086/670067}, 125, 306,
  [\href{https://arxiv.org/abs/1202.3665}{1202.3665}].

\bibitem[\protect\citeauthoryear{Frandsen \& Petersen}{Frandsen \&
  Petersen}{2018}]{Frandsen:2018ftj}
Frandsen M.~T.,  Petersen J.,  2018,
  [\href{https://arxiv.org/abs/1805.10706}{1805.10706}].

\bibitem[\protect\citeauthoryear{Gentile, Famaey  \& de Blok}{Gentile
  et~al.}{2011}]{Gentile:2010xt}
Gentile G.,  Famaey B.,   de Blok W.,  2011, \mn@doi [\aap]
  {10.1051/0004-6361/201015283}, 527, A76,
  [\href{https://arxiv.org/abs/1011.4148}{1011.4148}].

\bibitem[\protect\citeauthoryear{Gregory}{Gregory}{2010}]{gregory2010bayesian}
Gregory P.~C.,  2010, Bayesian logical data analysis for the physical sciences
  : a comparative approach with Mathematica support.
Cambridge University Press, Cambridge New York.

\bibitem[\protect\citeauthoryear{Jeffreys}{Jeffreys}{2011}]{9781447494782}
Jeffreys H.,  2011, Scientific Inference.
Muller Press.

\bibitem[\protect\citeauthoryear{Kent}{Kent}{1987}]{Kent:1987zz}
Kent S.~M.,  1987, \mn@doi [Astron. J.] {10.1086/114366}, 93, 816.

\bibitem[\protect\citeauthoryear{Kroupa et~al.,}{Kroupa
  et~al.}{2018}]{Kroupa:2018kgv}
Kroupa P.,  et~al., 2018, \mn@doi [\natastron] {10.1038/s41550-018-0622-x}, 2,
  925, [\href{https://arxiv.org/abs/1811.11754}{1811.11754}].

\bibitem[\protect\citeauthoryear{{Lelli}, {McGaugh}  \& {Schombert}}{{Lelli}
  et~al.}{2016}]{2016AJ....152..157L}
{Lelli} F.,  {McGaugh} S.~S.,   {Schombert} J.~M.,  2016, \mn@doi [\aj]
  {10.3847/0004-6256/152/6/157}, 152, 157,
  [\href{https://arxiv.org/abs/1606.09251}{1606.09251}].

\bibitem[\protect\citeauthoryear{Lewis}{Lewis}{2019}]{Lewis:2019xzd}
Lewis A.,  2019, [\href{https://arxiv.org/abs/1910.13970}{1910.13970}].

\bibitem[\protect\citeauthoryear{Li, Lelli, McGaugh  \& Schormbert}{Li
  et~al.}{2018}]{Li:2018tdo}
Li P.,  Lelli F.,  McGaugh S.,   Schormbert J.,  2018, \mn@doi [Astron.
  Astrophys.] {10.1051/0004-6361/201732547}, 615, A3,
  [\href{https://arxiv.org/abs/1803.00022}{1803.00022}].

\bibitem[\protect\citeauthoryear{Lin \& Ishak}{Lin \&
  Ishak}{2017}]{Lin:2017ikq}
Lin W.,  Ishak M.,  2017, \mn@doi [Phys. Rev.] {10.1103/PhysRevD.96.023532},
  D96, 023532, [\href{https://arxiv.org/abs/1705.05303}{1705.05303}].

\bibitem[\protect\citeauthoryear{Ludlow et~al.}{Ludlow
  et~al.}{2017}]{Ludlow:2016qzh}
Ludlow A.~D.,  et~al., 2017, \mn@doi [Phys. Rev. Lett.]
  {10.1103/PhysRevLett.118.161103}, 118, 161103,
  [\href{https://arxiv.org/abs/1610.07663}{1610.07663}].

\bibitem[\protect\citeauthoryear{McGaugh}{McGaugh}{2004}]{McGaugh:2004aw}
McGaugh S.~S.,  2004, \mn@doi [Astrophys. J.] {10.1086/421338}, 609, 652,
  [\href{https://arxiv.org/abs/astro-ph/0403610}{astro-ph/0403610}].

\bibitem[\protect\citeauthoryear{McGaugh, Lelli  \& Schombert}{McGaugh
  et~al.}{2016}]{McGaugh:2016leg}
McGaugh S.,  Lelli F.,   Schombert J.,  2016, \mn@doi [Phys. Rev. Lett.]
  {10.1103/PhysRevLett.117.201101}, 117, 201101,
  [\href{https://arxiv.org/abs/1609.05917}{1609.05917}].

\bibitem[\protect\citeauthoryear{McGaugh, Li, Lelli  \& Schombert}{McGaugh
  et~al.}{2018}]{McGaugh:2018aa}
McGaugh S.~S.,  Li P.,  Lelli F.,   Schombert J.~M.,  2018, \mn@doi
  [\natastron] {10.1038/s41550-018-0615-9}.

\bibitem[\protect\citeauthoryear{Meidt et~al.}{Meidt
  et~al.}{2014}]{Meidt:2014mqa}
Meidt S.~E.,  et~al., 2014, \mn@doi [\apj] {10.1088/0004-637X/788/2/144}, 788,
  144, [\href{https://arxiv.org/abs/1402.5210}{1402.5210}].

\bibitem[\protect\citeauthoryear{{Milgrom}}{{Milgrom}}{1983}]{1983ApJ...270..371M}
{Milgrom} M.,  1983, \mn@doi [\apj] {10.1086/161131}, 270, 371.

\bibitem[\protect\citeauthoryear{{Milgrom}}{{Milgrom}}{1988}]{1988ApJ...333..689M}
{Milgrom} M.,  1988, \mn@doi [\apj] {10.1086/166777}, 333, 689.

\bibitem[\protect\citeauthoryear{Milgrom}{Milgrom}{2015}]{Milgrom:2015ema}
Milgrom M.,  2015, \mn@doi [Phys. Rev.] {10.1103/PhysRevD.92.044014}, D92,
  044014, [\href{https://arxiv.org/abs/1507.05741}{1507.05741}].

\bibitem[\protect\citeauthoryear{Milgrom}{Milgrom}{2016}]{Milgrom:2016uye}
Milgrom M.,  2016, [\href{https://arxiv.org/abs/1609.06642}{1609.06642}].

\bibitem[\protect\citeauthoryear{Mo, van~den Bosch  \& White}{Mo
  et~al.}{2010}]{0521857937}
Mo H.,  van~den Bosch F.,   White S.,  2010, {Galaxy Formation and Evolution}.
Cambridge University Press.

\bibitem[\protect\citeauthoryear{Navarro, Ben{\'\i}tez-Llambay, Fattahi, Frenk,
  Ludlow, Oman, Schaller  \& Theuns}{Navarro et~al.}{2017}]{Navarro:2016bfs}
Navarro J.~F.,  Ben{\'\i}tez-Llambay A.,  Fattahi A.,  Frenk C.~S.,  Ludlow
  A.~D.,  Oman K.~A.,  Schaller M.,   Theuns T.,  2017, \mn@doi [\mnras]
  {10.1093/mnras/stx1705}, 471, 1841,
  [\href{https://arxiv.org/abs/1612.06329}{1612.06329}].

\bibitem[\protect\citeauthoryear{Profumo}{Profumo}{2017}]{9781786340016}
Profumo S.,  2017, Introduction To Particle Dark Matter, An (Advanced Textbooks
  in Physics).
WSPC (EUROPE).

\bibitem[\protect\citeauthoryear{{Querejeta} \& {et al}}{{Querejeta} \& {et
  al}}{2015}]{2015ApJS..219....5Q}
{Querejeta} M.,  {et al} 2015, \mn@doi [\apjs] {10.1088/0067-0049/219/1/5},
  219, 5, [\href{https://arxiv.org/abs/1410.0009}{1410.0009}].

\bibitem[\protect\citeauthoryear{Randriamampandry \& Carignan}{Randriamampandry
  \& Carignan}{2014}]{Randriamampandry:2014eoa}
Randriamampandry T.,  Carignan C.,  2014, \mn@doi [\mnras]
  {10.1093/mnras/stu100}, 439, 2132,
  [\href{https://arxiv.org/abs/1401.5619}{1401.5619}].

\bibitem[\protect\citeauthoryear{Ren, Kwa, Kaplinghat  \& Yu}{Ren
  et~al.}{2019}]{Ren:2018jpt}
Ren T.,  Kwa A.,  Kaplinghat M.,   Yu H.-B.,  2019, \mn@doi [Phys. Rev.]
  {10.1103/PhysRevX.9.031020}, X9, 031020,
  [\href{https://arxiv.org/abs/1808.05695}{1808.05695}].

\bibitem[\protect\citeauthoryear{Rodrigues, Marra, Del~Popolo  \&
  Davari}{Rodrigues et~al.}{2018a}]{Rodrigues:2018duc}
Rodrigues D.~C.,  Marra V.,  Del~Popolo A.,   Davari Z.,  2018a, \mn@doi
  [\natastron] {10.1038/s41550-018-0498-9}, 2, 668,
  [\href{https://arxiv.org/abs/1806.06803}{1806.06803}].

\bibitem[\protect\citeauthoryear{Rodrigues, Marra, Del~Popolo  \&
  Davari}{Rodrigues et~al.}{2018b}]{Rodrigues:2018lvw}
Rodrigues D.~C.,  Marra V.,  Del~Popolo A.,   Davari Z.,  2018b, \mn@doi
  [\natastron] {10.1038/s41550-018-0614-x}, 2, 927,
  [\href{https://arxiv.org/abs/1811.05882}{1811.05882}].

\bibitem[\protect\citeauthoryear{Rodrigues, Marra, Del~Popolo  \&
  Davari}{Rodrigues et~al.}{2020}]{Rodrigues:2020squ}
Rodrigues D.~C.,  Marra V.,  Del~Popolo A.,   Davari Z.,  2020, \mn@doi
  [\natastron] {10.1038/s41550-019-0999-1}, 4, 134,
  [\href{https://arxiv.org/abs/2002.01970}{2002.01970}].

\bibitem[\protect\citeauthoryear{{Sanders}}{{Sanders}}{1990}]{1990A&ARv...2....1S}
{Sanders} R.~H.,  1990, \mn@doi [\aapr] {10.1007/BF00873540}, 2, 1.

\bibitem[\protect\citeauthoryear{{Stone} \& {Courteau}}{{Stone} \&
  {Courteau}}{2019}]{2019ApJ...882....6S}
{Stone} C.,  {Courteau} S.,  2019, \mn@doi [\apj] {10.3847/1538-4357/ab3126},
  882, 6, [\href{https://arxiv.org/abs/1908.06105}{1908.06105}].

\bibitem[\protect\citeauthoryear{Verde, Protopapas  \& Jimenez}{Verde
  et~al.}{2013}]{Verde:2013wza}
Verde L.,  Protopapas P.,   Jimenez R.,  2013, \mn@doi [Phys. Dark Univ.]
  {10.1016/j.dark.2013.09.002}, 2, 166,
  [\href{https://arxiv.org/abs/1306.6766}{1306.6766}].

\bibitem[\protect\citeauthoryear{Zhou, Del~Popolo  \& Chang}{Zhou
  et~al.}{2020}]{Zhou:2020tst}
Zhou Y.,  Del~Popolo A.,   Chang Z.,  2020, \mn@doi [Phys. Dark Univ.]
  {10.1016/j.dark.2020.100468}, 28, 100468.

\bibitem[\protect\citeauthoryear{de Almeida, Amendola  \& Niro}{de~Almeida
  et~al.}{2018}]{deAlmeida:2018kwq}
de Almeida {\'A}. O.~F.,  Amendola L.,   Niro V.,  2018, \mn@doi [JCAP]
  {10.1088/1475-7516/2018/08/012}, 1808, 012,
  [\href{https://arxiv.org/abs/1805.11067}{1805.11067}].

\bibitem[\protect\citeauthoryear{{de Blok} \& {McGaugh}}{{de Blok} \&
  {McGaugh}}{1998}]{1998ApJ...508..132D}
{de Blok} W.~J.~G.,  {McGaugh} S.~S.,  1998, \mn@doi [\apj] {10.1086/306390},
  508, 132, [\href{https://arxiv.org/abs/astro-ph/9805120}{astro-ph/9805120}].

\makeatother
\end{thebibliography}


\appendix

\section{$\mathbf{5\sigma}$ credible intervals} \label{app:mcmc}

\begin{figure*}
\centering
\includegraphics[width=\textwidth]{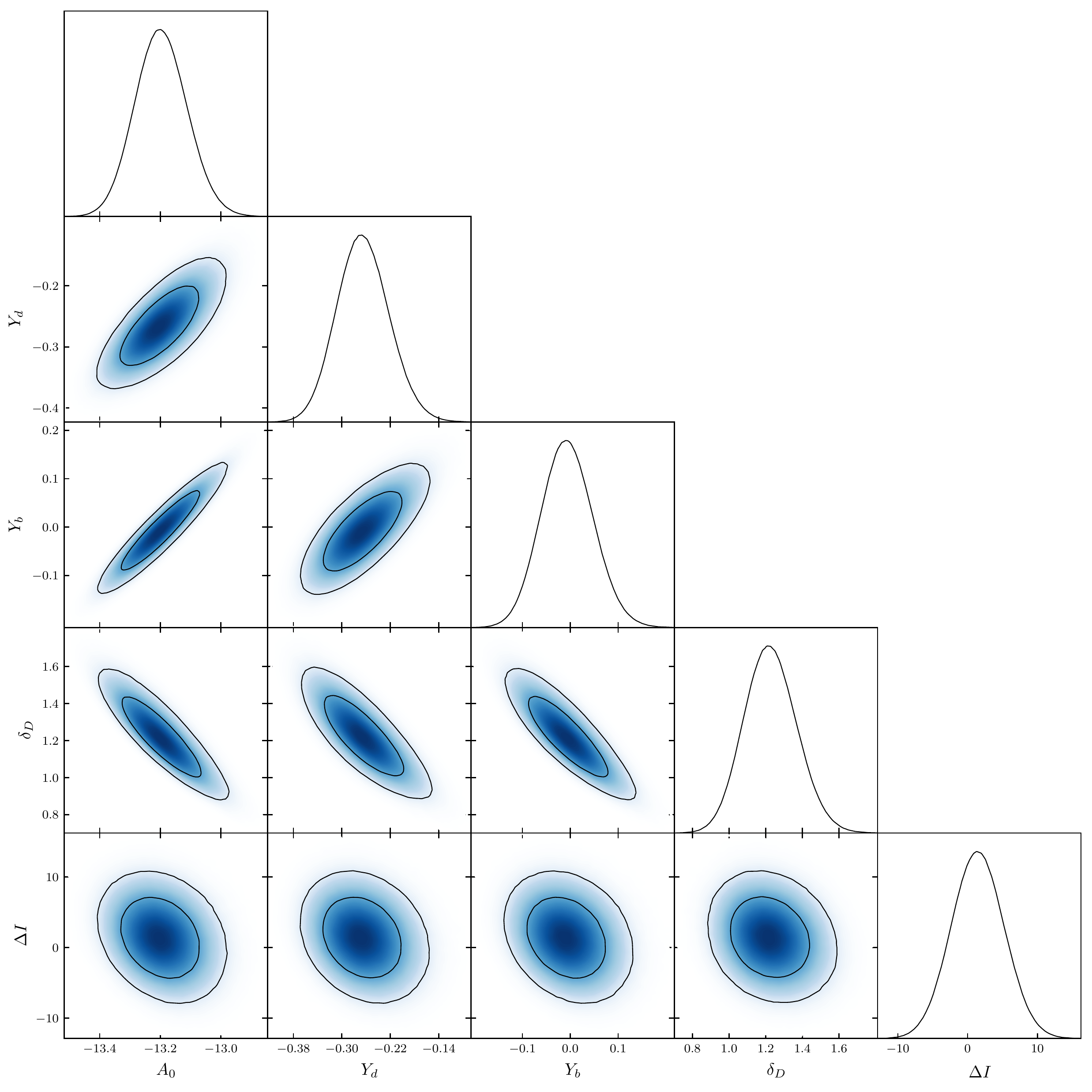}
\caption{Marginalized 1 and 2$\sigma$ contours for the galaxy UGC 03205 of the SPARC dataset.
Here, $\delta_D=D/D_0$ and $\Delta I = I-I_0$.}
\label{fig:triplot}
\end{figure*}

\begin{figure}
\centering
\includegraphics[width= \columnwidth]{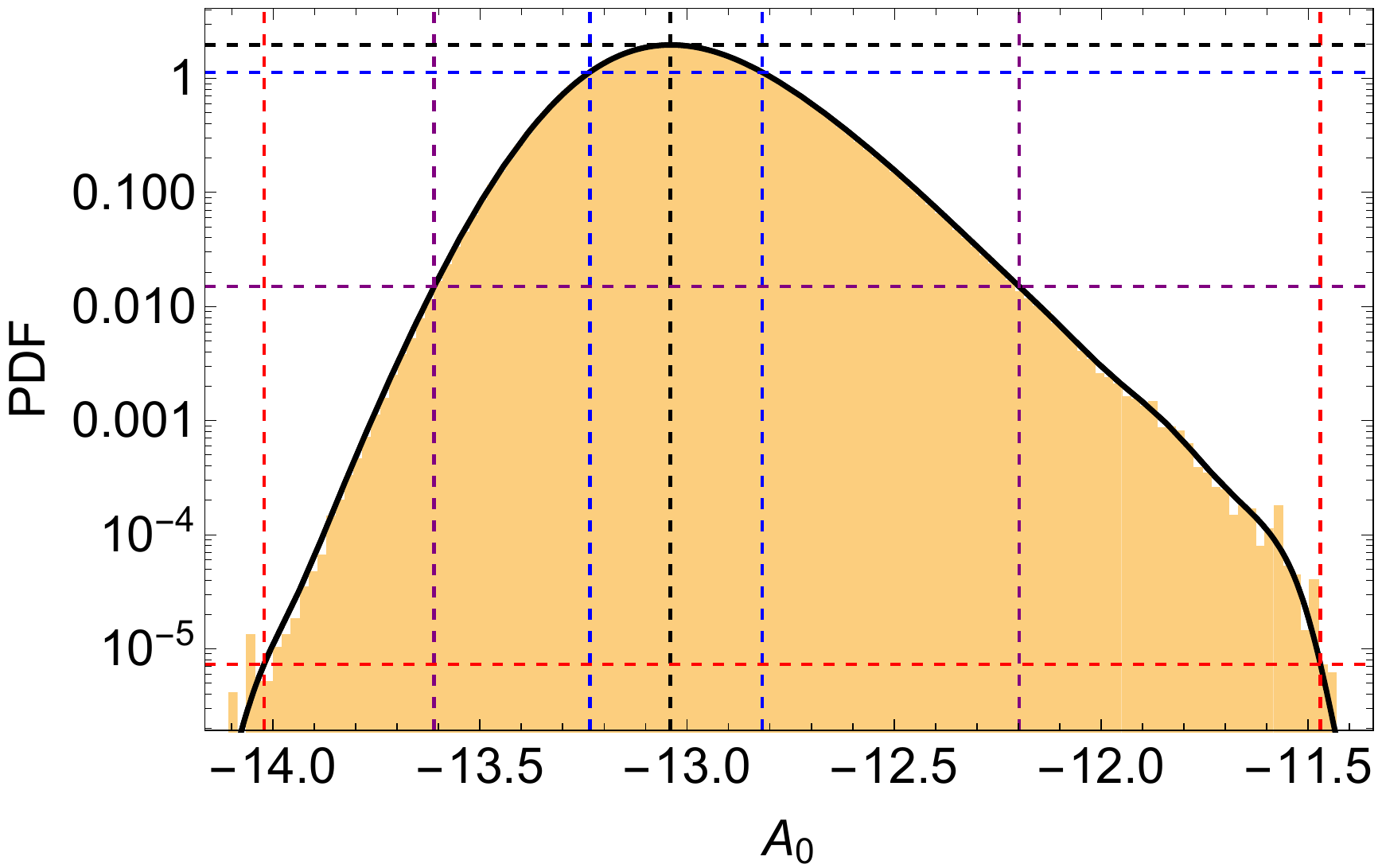}
\caption{Marginalized MCMC histogram (orange shade) and corresponding smoothed marginalized posterior $f(\loga| \text{F574-1})$ (black line). The dashed black horizontal and vertical lines mark the mode. The blue, purple and red lines mark the 1, 3 and 5$\sigma$ credible intervals, respectively.}
\label{fig:levels}
\end{figure}

As an example, in Figure~\ref{fig:triplot} we show the triangular plot that illustrates the MCMC exploration of the posterior $f(\theta| \text{UGC 03205})$.
As another example, in Figure~\ref{fig:levels} we show the marginalized MCMC histogram (orange shade) and the corresponding smoothed marginalized posterior $f(\loga| \text{F574-1})$ (black line). The fact that the MCMC points span the 5$\sigma$ interval shows that the adopted number of $45 \times 10^{6}$ points is adequate.

\section{Posterior distribution extension beyond 5$\sigma$} \label{app:ext}

In order to proceed as described in Sections~\ref{analysis} one needs to have posterior distributions $f(\loga| {\rm galaxy}_{\!g})$ defined in the common prior interval $-20 \le \loga \le -5$.
We obtained the posteriors using a very large number of MCMC points ($45 \times 10^{6}$), which is enough to reliably obtain 5$\sigma$ credible intervals. These posteriors, however, do not span the interval $-20 \le \loga \le -5$ as it is not numerically viable to reconstruct the posteriors up to, e.g., $>10\sigma$.
In order to overcome this technical difficulty, we extended the distributions $f(\loga| {\rm galaxy}_{\!g})$ beyond their $5\sigma$ intervals using the following Gaussian tails as defined in eq.~\eqref{eqB1}. In that equation, $const\simeq 1$ is a normalization constant, $A_0^{\rm min} \le \loga \le A_0^{\rm max}$ is the region covered by the MCMC chain, and $\loga_{g}^\text{mean}$ and $\sigma_{g}$ are mean and standard deviation relative to $f(\loga| {\rm galaxy}_{\!g})$.

\begin{table*}
	\begin{align}
	&f^{\rm ext}(\loga | {\rm galaxy}_{\!g}) = const \times \left \lbrace
	\renewcommand{\arraystretch}{2}
		\begin{array}{ll}
		f(A_0^{\rm min}| {\rm galaxy}_{\!g}) \exp\left[{\frac{(A_0^{\rm min} - A_{0 g}^\text{mean})^2-(\loga - A_{0g}^\text{mean})^2}{2 \sigma_g^{2}}}\right]  & \; -20 \le \loga \le A_0^{\rm min} \\
		f(\loga| {\rm galaxy}_{\!g})  & \; A_0^{\rm min} \le \loga \le A_0^{\rm max}  \label{eqB1}\\
		f(A_0^{\rm max}| {\rm galaxy}_{\!g})  \exp \left[{\frac{(A_0^{\rm max} - A_{0 g}^\text{mean})^2-(\loga - A_{0g}^\text{mean})^2}{2 \sigma_{g}^{2}}} \right]  & \;  A_0^{\rm max} \le \loga \le -5
		\end{array}  \right. 
	\end{align}
\end{table*}

\section{Review on the Gaussian approximation method} \label{app:revGau}
For completeness, and since it will be useful for comparing with our newer results, we briefly review below the Gaussian approximation and $\chi^2$ statistics that we used in \Rteen{}.

In \citetalias{Rodrigues:2018duc} we computed the global best value by approximating the posteriors $f(\loga| {\rm galaxy}_{\!g})$ as Gaussians, that is, we neglected higher moments,
\begin{align}
f(\loga| {\rm galaxy}_{\!g}) \approx f_{\rm gau}(\loga| {\rm galaxy}_{\!g}) \equiv \mathcal{N}(\loga_{g}^\text{mean}, \sigma_{g}) \,,
\end{align}
where $\loga_{g}^\text{mean}$ and $\sigma_{g}$ are mean and standard deviation relative to $f(\loga| {\rm galaxy}_{\!g})$.
Maximizing $\prod_g f_{\rm gau}(\loga| {\rm galaxy}_{\!g})$ is equivalent to minimizing the following $\chi^2$ function (as done in \citetalias{Rodrigues:2018duc}):
\begin{equation} \label{chi2a0}
\chi^{2}(\loga) = \sum_{g=1}^{N_{\rm G}} \frac{(\loga_{g}^\text{mean}-\loga)^{2}}{\sigma_{g}^{2}} \,,
\end{equation}
so that $\chi^{2}(A_0^{\rm gbv})=\chi^{2}_{\rm min}$.


\bsp	
\label{lastpage}
\end{document}